\documentstyle[aps,prl,amssymb,float,graphicx,twocolumn]{revtex}
\begin{document}
\draft \wideabs{
\title{Critical current in charge-density wave transport}
\author{A.~A.~Sinchenko}
\address{Moscow state  Engineering-Physics Institute, 115409 Moscow, Russia}
\author{V.~Ya.~Pokrovski, S.~G.~Zybtsev, I.~G.~Gorlova, Yu.~I.~Latyshev}
\address{Institute of Radioengineering and Electronics, Russian Academy of
Sciences, 103907 Moscow, Russia}
\author{P.~Monceau}
\address{Centre de Recherches sur Les Tr\`{e}s Basses
Temp\'{e}ratures, C.N.R.S., B.P. 166, 38042, Grenoble C\'{e}dex 9,
France}
\date{\today}
\maketitle
\begin{abstract}
We report transport measurements under very high current densities
$j$, up to $\sim10^8$~A/cm$^2$, of quasi-one-dimensional
charge-density wave (CDW) conductors NbSe$_3$ and TaS$_3$.  Joule
heating has been minimized by using a point-contact configuration
or by measuring samples with extremely small cross-sections. Above
$j_c \approx 10^7$~A/cm$^2$ we find evidence for suppression of
the Peierls gap and development of the metallic state. The
critical CDW velocity corresponding with $j_0$ is comparable with
the sound velocity, and with $\Delta/ \hbar k_F$ ($k_F$ is the
Fermi wave vector), which corresponds to the depairing current.
Possible scenarios of the Peierls state destruction are discussed.
\end{abstract}

}

The discovery of non-Ohmic conductivity in  quasi-one-dimensional
charge-density wave (CDW) materials has opened great expectations
concerning the so-called Fr\"ohlich superconductivity \cite{fro}.
However, the experiments up to now have showed that the CDW
conductivity under the highest electric fields asymptotically
approaches a value close to the "normal-state conductivity" and
never exceeds it. This conductivity is estimated from the
extrapolation of the temperature dependent conductivity above the
Peierls transition \cite{Grun}. The hypothesis which can be
naturally risen, namely that the CDW state is suppressed at high
currents was rejected by x-rays measurements of the CDW satellite
profiles in the sliding state \cite{fle},\cite{ring} and by
narrow-band noise measurements at high current densities $j$ (up to
$3\cdot 10^4$~A/cm$^2$ for TaS$_3$ \cite{Mon}): in the latter case,
it was shown that the main part of the current is carried by the
coherently sliding CDW with no reduction of its charge density. On
the other hand, the CDW velocity cannot grow up to infinity. So,
the question concerning the existence of an upper limit for the
velocity of the sliding CDW is still undecided up to now.

To our knowledge, it has been reported only one indication in
favour of the local suppression of the Peierls energy gap
$2\Delta$, and that in K$_{0.3}$MoO$_3$ \cite{SinchJETP}. This
effect, attributed to a high current density near the metal -- CDW
surface  was observed at a high degree of injection of normal
carriers through a Au--K$_{0.3}$MoO$_3$ boundary. Estimation of
the critical current density gave a value of $j_c=4.8\cdot10^7$
A/cm$^2$. However, it is not clear if the gap suppression reported
in Ref.\onlinecite{SinchJETP} is associated with the high CDW
velocity, or  is induced by the injected quasi-particules. For a
clarification of the situation it would be highly desirable to
observe this phenomenon directly. Evidently, Joule heating in the
previous experimental configurations will not allow to achieve
high enough values of $j$.

Hereafter we propose two approaches for overcoming this problem. The first
one is the point contact configuration formed between two thin
whiskers of NbSe$_3$. It is well known that the electric field is
localized near a point contact in a small region with a
characteristic size of the order of the point contact diameter,
$d$. In the case $d<<l$ ($l$ is the mean free path) the length for energy
relaxation is much longer than that corresponding to the formation of
the point contact resistance  and the heating effect is strongly suppressed \cite{Kulik}.
Thus, a point contact is a very convenient configuration for high current
measurements. We have used the
configuration proposed in Ref.~\onlinecite{SinchPRB}. The electric
contact between two NbSe$_3$ stripe--like single crystals with
perpendicular $bc$ -- planes is formed at low temperatures by
means of a precise mechanical motion transfer system, so that
the current through the contacts flows along the $b$-axis. The
samples selected for the experiment had typical dimensions: along
the $b$-axes $L_b\approx 1$~mm, along the $c$-axes $L_c\approx
10\div 50$~$\mu$m and along the $a^*$- axes $L_a\approx 1$ $\mu$m.

Disadvantages of the point contact configuration are a probability
for having a non-predicted barrier at the boundary between the
samples and a non-uniform current distribution. So we have
performed high--current measurements of extremely thin samples of
TaS$_3$ with a cross-section area $\lesssim 10^{-3}$~$\mu$m$^2$
and a typical length $5 \div 10$~$\mu$m. The cross-section area is
estimated from the value of the room-temperature resistance of the
samples \cite{BZN}. Due to the very good thermal contact with the
sapphire substrates such samples can sustain extremely high
current densities with a relatively small Joule heating. The main
results presented below are insensitive to the exact value of
heating.
\begin{figure}[!t]
\includegraphics[scale=.5]{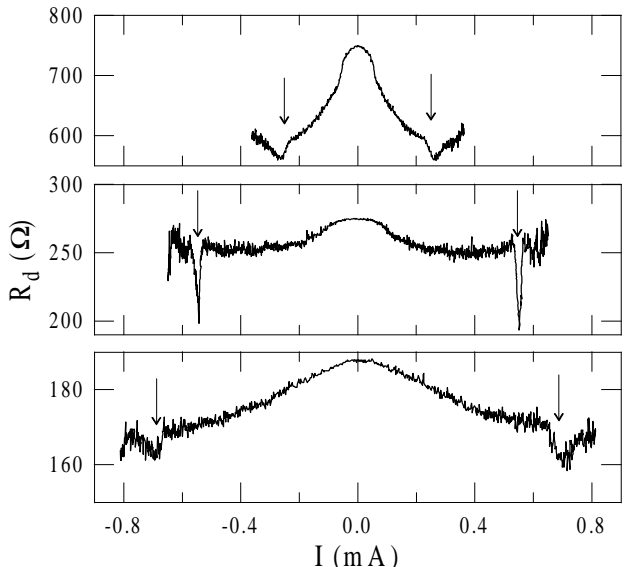}
\caption{Current dependence of the differential resistance $R_d(I)$
of NbSe$_3$--NbSe$_3$ point contacts. $T$~= 77~K. The arrows
indicate $I_0$} \label{Fig1}
\end{figure}
Let first analyse our results with the point contact configuration.
Figure 1 shows typical dependencies of the differential resistance
$R_d(I)$ for three different NbSe$_3$-NbSe$_3$ point contacts
obtained at $T$=77~K.  $R_d(I)$ monotonously decreases with the
increase of the injected current revealing the CDW sliding. The
threshold electric field for initiation of the CDW sliding, $E_T$,
is achieved at a very small current, so the plateau in the curves
corresponding to the pinned state of the CDW is indistinguishable
in this scale. At very high current $R_d$ is close to the
resistance values which would be observed in the absence of the
Peierls transition \cite{Grun}. In this region, practically for all
investigated contacts we clearly observed a sharp decrease of the
resistance at the current $I_0$ (indicated by arrows in Fig.1).
Above $I_0$ the resistance is slightly growing with current,
evidently reflecting the Joule heating. The growth of $R$ with
heating should be the case for a metallic state. We could not
measure the $R_d(I)$ dependence far above $I_0$ because the
contacts are very unstable in this region and often burn.

The exact determination of the CDW current density corresponding to
$I_0$ is complicated in the case of NbSe$_3$. The unit cell of
NbSe$_3$ contains three different types of chains. In the
temperature range from 59~K to 145~K two types of chains are in the
normal state and one is in the CDW state \cite{Grun}, but at high
electric field we can assume that the injected current is
homogeneously distributed over the chains, because in this case the
CDW conductivity is close to the normal-state value. So $I\approx
j\pi d^2/4$, where $j$ is the current density, and the point
contact diameter may be estimated from the well known Sharvin
formula \cite{Sharvin}: $R_{ds}=l\rho/d^2$, where $\rho$ is the
normal state resistivity and $R_{ds}$ is the saturation value of
$R_d$ at high current. Fig. 2 shows the variation of $I_0$ as a
function of $1/R_{ds}$, which is proportional to the point--contact
area, for more than 20 different point contacts. The dependence is
close to be linear. The proportionality between $I_0$ and the
conductivity means that the current density is constant for all
contacts and is evidently determined by fundamental properties of
the material. For $\rho=3\cdot10^{-4}$ $\Omega$cm (Ref
\onlinecite{OngBrill78}) and for $l$=100 nm, we estimate the mean
value for the CDW current density $j_0=5.9\cdot10^7$ A/cm$^2$.
\begin{figure}[!t]
\includegraphics[width=8cm,height=6cm]{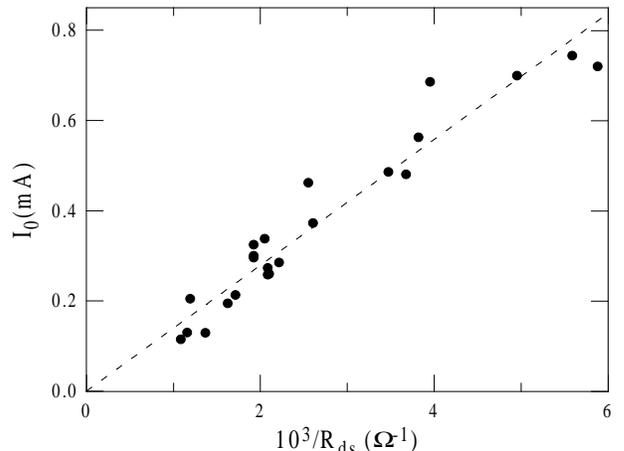}
\caption{Critical current $I_0$ as a function of inverse saturation
differential resistance of NbSe$_3$--NbSe$_3$ point contacts.}
\label{Fig2}
\end{figure}
For making clear if the abrupt change in conductivity of NbSe$_3$
at $j \sim 10^7$~A/cm$^2$ really reveals a transition into a
metallic state, i.e. a critical (depairing) current for the CDW,
or is associated with contact phenomena under inhomogeneous
conditions, we performed transport measurements of extremely fine
samples of TaS$_3$. We selected this material, as it demonstrates
a purely dielectric behavior below $T_P \approx 220$~K. So, a
transition into a metallic state should be more obvious for it.
Besides, TaS$_3$ crystals can be easily splitted and make possible
the preparation of samples with a cross section area below
$10^{-3}$~$\mu$m$^2$.

Typical results of transport measurements are shown in Figure 3,
which illustrates ``raw'' dependencies of the conductivity of a
representative sample at different temperatures as a function of
the current density (open symbols). These data have to be
corrected of Joule heating. We have estimated its value by two
ways. First, we noticed, that at temperatures well above $T_P$
(typically $T>250$~K) the non-linear conductivity due to CDW
fluctuations becomes negligible; therefore, the slight change of
resistance with voltage, $\delta R \equiv R(V)-R(0) \propto V^2$,
is only provided by heating. The comparison of $\delta R$ with
$dR/dT$ at several temperatures yields the estimate of heating as
a function of Joule power, $W$. The second estimate is based on
the dependencies $R(j)$ obtained at low temperatures. We noticed
that at the highest currents the temperature of the samples
exceeds $T_P$, and that at a certain current density each curve
$R(j)$ (and $R_d(j)$) demonstrates a minimum  corresponding with
the minimum of $R(T)$, which is observed around $T_M = 250 \div
300$~K. Thus, the power $W_M$ at the minimum of $R(j)$ provides
the estimate of heating $\delta T=T_M-T$, where $T$ is the ambient
temperature. The dependence $W_M$ {\it vs.} $T$ is approximately
linear, its slope giving the required value $dT/dW$. Both ways
give $dT/dW \approx 10^5$~K/W for samples with dimensions
$5$~$\mu$m $\times 10^{-3}$~$\mu$m$^2$. Knowing the value of $W$
at each $V$ and $dT/dW$ we thus know the temperature for each
point of each curve. Interpolating the set of I-V curves we got
isothermal dependencies and thus the correction for heating. The
curves obtained after this temperature correction are shown in
Fig.3 (solid lines). One can see that the curves for $T<T_P$
intersect at $j=10^7$~A/cm$^2$. For $j > 10^7$~A/cm$^2$ the growth
of conductivity tends to saturate value. However, we cannot
determine exactly the asymptotic behavior of the curves, because
our results for high currents  are very sensitive to the value of
$dT/dW$ used for the temperature correction.
\begin{figure}[!t]
\includegraphics[width=8cm,height=5.6cm]{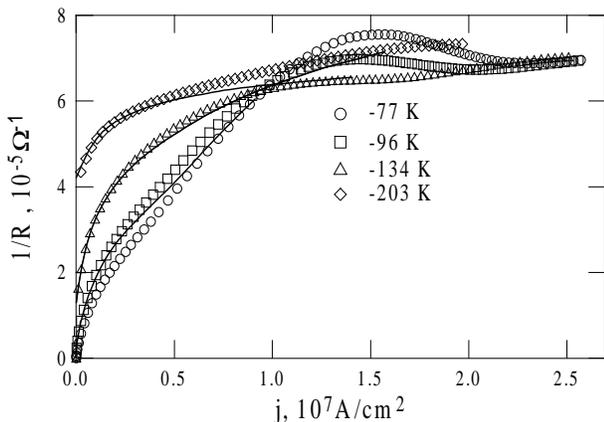}
\caption{Conductivity as a function of $j$ at several temperatures
indicated in the plot. The solid lines show data after the
temperature correction with $dT/dW=1.3 \cdot 10^5$~K/W. The
dimensions of the TaS$_3$ sample are $4.7$~$\mu$m$\times
10^{-3}$~$\mu$m$^2$.} \label{Fig3}
\end{figure}
To ascertain if TaS$_3$ really exhibits a transition into a
metallic state, we have plotted the conductivity measured at a
given current as a function of temperature. Fig.4 shows such a set
of dependencies. For comparison we also show the temperature
dependence of the conductivity at a low field $E \ll E_T$ ($E_T
\approx 25$~V/cm). For $E\lesssim 3$~kV/cm the temperature
dependence of the conductivity demonstrates a dielectric behaviour
($dR/dT<0$). While for $E \lesssim 500$~V/cm, the high-field
conductivity approximately follows the behaviour of the
conductivity measured at low-field  \cite{scaling}, at higher $E$
the activation energy decreases, and for $E_0=3.3$~kV/cm
($j_0=10^7$~A/cm$^2$) the conductivity is nearly independent of
temperature. Increasing the current further results in the metallic
behaviour of conductivity, {\it i.e.} $dR/dT>0$. We emphasize that
this result and the value of the current $j_0$ are quite
insensitive to the value of $dT/dW$ taken for the Joule heat
correction \cite{10K}.
\begin{figure}[t]
\includegraphics[width=8cm,height=7.5cm]{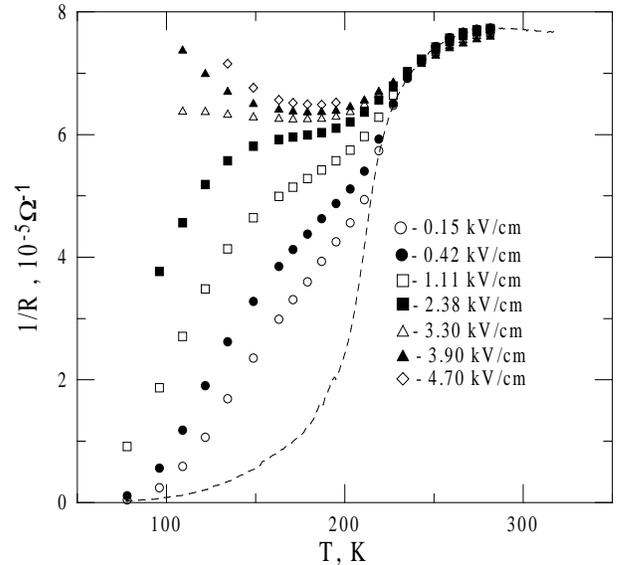}
\caption{A set of temperature dependencies of the conductivity of
the representative TaS$_3$ sample at fixed electric fields
indicated in the plot. The broken line shows the conductivity at $E
\ll E_T$ ($E=0.1$~V/cm).} \label{Fig 4}
\end{figure}
The value of the non-linear contribution to $j_0$ as a function of
temperature is presented in Fig.5 for TaS$_3$ (black circles) and
for a stable point-contact NbSe$_3$-NbSe$_3$ for which we
succeeded in obtaining the temperature dependence of the current
$I_0$ (open circles). Both temperature and current scales are
normalized by the value of $T_p$ (145 K for NbSe$_3$ and 208 K for
TaS$_3$) and the value of $j_0$ at $T=0.62T_p$ respectively. As
seen in the figure, the value of $j_0$ for both materials has a
little tendency to decrease with increasing temperature. The
growth of $j_0$ for TaS$_3$ at $T \to T_P$ is probably due to the
growth of non-collective contributions to the current. In fact, at
the highest temperatures (Fig.5) the linear contribution to $j_0$,
{\it i.e.} $j_0*R/R(0)$),  is comparable with the non-linear part
of the current.
\begin{figure}[b]
\includegraphics[width=8cm,height=6cm]{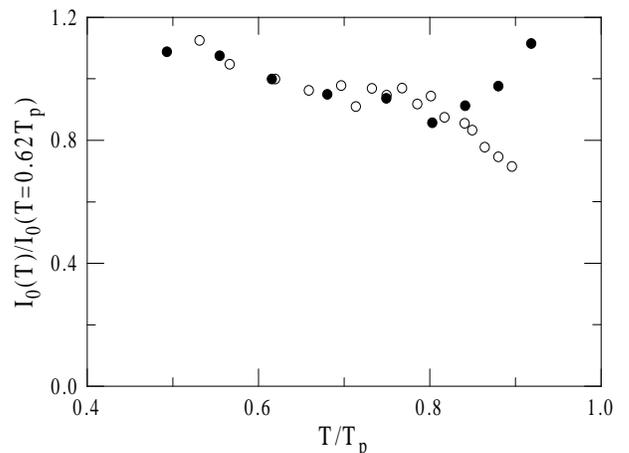}
\caption{Normalized temperature dependencies of critical current
densities for TaS$_3$ (black circles) and NbSe$_3$ (open circles).
The linear contribution to $j$ is subtracted} \label{Fig5}
\end{figure}

Analyzing our experimental results with the point contact technique
we are led to the assumption that the observed anomalies at $j=j_0$
are related to the suppression of the CDW state at high current
density. Indeed, the conductivity of the CDW is less than the
normal state conductivity at any current. So, the sharp drop of the
resistance of the NbSe$_3$--NbSe$_3$ point contacts reveals the
transition from the CDW conductivity to the normal state
conductivity. Similarly, a metallic behaviour is demonstrated in
very small cross-section TaS$_3$ samples above $j_0$. The fact that
the transition into the metallic state in the latter case is more
gradual can be ascribed to the non-uniform distribution of the CDW
velocities in the volume of the sample, while the point contact
probes only several wavelengths. Evidently, the transition of
TaS$_3$ into a metallic state is not complete, and in the vicinity
of $T=T_p$ the dielectric behaviour is observed up to the highest
fields (Fig.5). By analogy with superconductors this may be
associated with the development of a kind of mixed state in the CDW
at high currents.

The magnitude of the critical current densities for different CDW
materials are approximately the same: $j_0=5.9\cdot10^7$ A/cm$^2$
for NbSe$_3$, $10^7$ A/cm$^2$ for TaS$_3$ and $4.8\cdot10^7$
A/cm$^2$ for K$_{0.3}$MoO$_3$ \cite{SinchJETP}. It has to be noted
that, for the latter compound, the suppression of the gap has been
observed directly.

What could be the physical mechanism of this transition? The
theoretical description or even the consideration of this
phenomenon  does not exist up to now. Therefore, we can only
propose qualitative explanations of the effects we have observed.
First of all, let estimate the velocity of the CDW motion
corresponding to the critical current density, $j_0$. The usual
formula $v=j_0/ne$, where $n$ is the density of condensed carriers,
yields $v=2.7\cdot10^5$ cm/s for NbSe$_3$; $0.2\cdot10^5$ cm/s for
TaS$_3$ and $0.6\cdot10^5$ cm/s for K$_{0.3}$MoO$_3$, which are
close to the sound velocities, $v_s$, in these materials \cite{vs}.
As the CDW results from an electron-phonon interaction, 
%the CDW
%sliding is accompanied by an acoustic wave with the same velocity.
%Therefore, 
it is unclear how the CDW could survive at $v>v_s$  .
So, anomalies in the Peierls gap behaviour, presumably its
suppression, is expected at $v\sim v_s$.

The critical current can also reveal the electron-hole depairing,
by analogy with superconductors. In the sliding state, the Fermi
distribution $\varepsilon(k)$ is shifted by a quantity $\delta k=k_F v/v_F$
and distorted by $\delta \varepsilon=\pm v\hbar q/2$ at $k= \pm
k_F$, where $q=2k_F$ is the CDW wave vector \cite{Grun,Al74}. One
may consider that the CDW gap will be suppressed when the
distortion  $\delta \varepsilon$ will
reach a value comparable to $\Delta$: $v\hbar q \sim \Delta$. This
gives the estimate of depairing current density $\lesssim
10^8$~A/cm$^2$. The same estimate arises from quite another
consideration. Note, that the frequency of the narrow-band noise
generation (the so-called fundamental frequency) $f\approx vq/2\pi$
for the depairing current density is $\sim \Delta/h$. Suppose, at some
point (say at a pinning center) the gap is suppressed, {\it i.e.} a
phase slip (PS) act occurs. From the principle of uncertainty, the
lowest possible time for the gap suppression or recovering is $\sim
h/\Delta$, which could be treated as the lowest possible time of a
PS act. If the PS frequency exceeds the critical value, the gap has
no time to restore before another wavelength comes, and the area of
the gap suppression expands throughout the sample. So, the PS time
limitation from below gives the same value of critical current.
Nucleation of the metallic state resulting from a continuous
increase of PS processes was considered earlier as an approach to
the Peierls transition from low temperatures \cite{MbI}.

In conclusion, we have for the first time considered the question
about the critical velocity of the CDW and proposed an experimental
answer to it. For TaS$_3$ and NbSe$_3$ we clearly observed a
Peierls state -- normal metal transition at $j_0\sim10^7$~A/cm$^2$.
The critical velocity, $v_0$, of the sliding CDW corresponding to
the value of $j_0$ approaches the speed of sound, and is comparable
with $\Delta/q$.

We are thankful to S.V.~Zaitsev-Zotov for permanent help in the
experiment and discussions, and to S.N.~Artemenko for helpful
discussions. This work has been supported by the Russian State
Foundation for Basic Research (grants No 99--02--17364, No
99--02-17387, and No 98--02--16667), by the twinning research
programme 19 from CNRS, and by the State programme "Physics of
Solid-State nanostructures" (No 97-1052).

\end{document}